\documentclass{aa}
\usepackage[pdftex]{graphicx}
\usepackage{txfonts}
\def \etal {\textit{et al. }}

\begin{document}
   \title{Searching For Transiting Circumbinary Planets in CoRoT \thanks{Based on observations obtained with CoRoT, a space project operated by the French Space Agency, CNES, with participation of the Science Programme of ESA, ESTEC/RSSD, Austria, Belgium, Brazil, Germany and Spain.} and Ground-Based Data Using CB-BLS}
   \author{A. Ofir \inst{1} \and H. J. Deeg \inst{2} \and C. H. S. Lacy \inst{3}}

   \institute{School of Physics and Astronomy, Raymond and Beverly Sackler Faculty of Exact Sciences, Tel Aviv University, Tel Aviv, Israel. \email{avivofir@wise.tau.ac.il}
		\and    Instituto de Astrofisica de Canarias, C. Via Lactea S/N, 38205 La Laguna, Tenerife, Spain
		\and Department of Physics, University of Arkansas, Fayetteville, AR 72701, USA
				}

   \date{Received XXX; accepted YYY}

% \abstract{}{}{}{}{} 
% 5 {} token are mandatory
 
  \abstract
  % context heading (optional)
  % {} leave it empty if necessary  
   {}
  % aims heading (mandatory)
   {Already from the initial discoveries of extrasolar planets it was apparent that their population and environments are far more diverse than initially postulated. Discovering circumbinary (CB) planets will have many implications, and in this context it will again substantially diversify the environments that produce and sustain planets. We search for transiting CB planets around eclipsing binaries (EBs).}
  % methods heading (mandatory)
   {CB-BLS is a recently-introduced algorithm for the detection of transiting CB planets around EBs. We describe progress in search sensitivity, generality and capability of CB-BLS, and detection tests of CB-BLS on simulated data. We also describe an analytical approach for the determination of CB-BLS detection limits, and a method for the correct detrending of intrinsically-variable stars.}
  % results heading (mandatory)
   {We present some blind-tests with simulated planets injected to real CoRoT data. The presented upgrades to CB-BLS allowed it to detect all the blind tests successfully, and these detections were in line with the detection limits analysis. We also correctly detrend bright eclipsing binaries from observations by the TrES planet search, and present some of the first results of applying CB-BLS to multiple real light curves from a wide-field survey.}
  % conclusions heading (optional), leave it empty if necessary 
   {CB-BLS is now mature enough for its application to real data, and the presented processing scheme will serve as the template for our future applications of CB-BLS to data from wide-field surveys such as CoRoT. Being able to put constraints even on non-detection will help to determine the correct frequency of CB planets, contributing to the understanding of planet formation in general. Still, searching for transiting CB planets is still a learning experience, similarly to the state of transiting planets around single stars only a few years ago. The recent rapid progress in this front, coupled with the exquisite quality of space-based photometry, allows to realistically expect that if transiting CB planets exist - then they will soon be found.}
   \keywords{methods: data analysis -- stars: variables: general -- stars: planetary systems -- occultations -- binaries: eclipsing}

\titlerunning{Searching For Transiting Circumbinary Planets Using CB-BLS}

\maketitle
%
%________________________________________________________________

% ========================
\section{Introduction}
% ========================

Bound binary stars are one of the most common environments in the Galaxy. Studying planet formation and evolution without accounting for binary (and multiple) star systems is surely incomplete -- so we aim to try to fill-in observational data on these types of systems. To help achieve this goal, a Binaries Working Group was recently formed within CoRoT Exoplanets Science Team of the CoRoT space mission (Baglin \etal 2006) to coordinate the different searches for planets in binary systems, and we report on one of the activities within the Working Group, namely -- transit searches using the CB-BLS algorithm (Ofir 2008, hereafter paper I). These efforts are now all the more relevant following the recent detection by Lee \etal (2009) of an eclipse-timing signal compatible with two circumbinary planets around HW Virginis.

Planets around eclipsing binaries may be expected in significant numbers, since CoRoT finds close and deeply eclipsing binaries in about 1\% of the stars surveyed (Almenara \etal 2009), leading to a sample of about 300 close binaries from its first year of operation. Assuming that a planetary orbit would be roughly aligned with a binary component's orbital plane, EBs constitute a sample with planetary orbits being preferentially aligned to display transits across the host stars.

There are many detection techniques for planets in binaries: Some are common to planets around single stars (such as: radial velocity (RV), transits, astrometry and microlensing [Muterspaugh \etal 2007, Lee \etal 2008]) and some are unique to binaries (such as: eclipse timing [Deeg \etal 2000, Lee \etal 2009] and gravitational waves [Seto 2008]). Generally speaking, in each geometry (i.e., circumprimary or circumbinary orbits) techniques that are in principle the same take on different emphasizes. 

In this paper we will focus on transits of CB planets, and specifically transiting CB planets around eclipsing binaries (EBs). The transit signal from CB planets is not periodic, hence common techniques searching such signals are of low efficiency and special search algorithms have to be employed, such as CB-BLS [paper I] or TDA [Doyle \etal 2000]. In the following we list the accumulated additions to CB-BLS since paper I in \S 2, describe blind tests to CB-BLS in \S 3, and show examples from the first application of CB-BLS to real data in \S 4, and conclude.

% ========================
\section{Accumulated Additions to CB-BLS}
\label{Additions}
% ========================

The following describe additions to the CB-BLS implementation starting from the version used to generate the results of paper I (internally designated as version 0.51) till now (version 0.83). While the basic idea of orbit fitting remains unchanged, the specific implementation has improved significantly in search sensitivity, generality, speed and results analysis. To avoid extensive repetitions, we assume in the following that the reader is familiar with paper I. We note that we plan to release the CB-BLS source code (written in \texttt{MATLAB}) in the future. CB-BLS now:

\begin{itemize}
\item  Allow the inclusion of the surface brightness ratio ($J$) in the input, which increases sensitivity. See more in \S \ref{OneDepth}.
\item  Allows natural use of the more accurate Roche-lobe geometry. All the user has to do is to specify the surface potentials $\Omega_{1,2}$ -- which are the direct outputs of EB modeling -- instead of the radii. CB-BLS then computes the 3D shape of the surface, rotates the shape to the binary inclination, and computes the silhouette of the sky-projected shape at each binary phase. CB-BLS allows single-surface binaries (contact binaries) too -- which are common. We note that this feature was forecasted in paper 1 (\S 3.2) and is implemented along these lines. We comment that using this feature slows CB-BLS significantly. 
\item  Accounts for EB inclination. The EB orbital inclination is calculated anyhow during the EB modeling, so the 2D (instead of 1D, as in paper I) sky-projected orbit for each component is computed - and the 2D distance from the test planetary model found. The planetary orbit is still assumed edge-on, but this is a good approximation since the planet is farther away from the baricenter than the EB components, and so it's alignment requirements are more stringent to begin with.
\item  Enables the simultaneous solution of multiple light curves of the same system, each with it's own EB model - which also allows one to use multi-band data in a single run. Since after regularization (see paper I \S 3.3) the transit signal should be achromatic - the CB-BLS statistic can be calculated using all the regularized residuals -- regardless of band. We note that performing a multi-band solution and using the (single) surface brightness ratio $J$ are mutually exclusive.
\item  Includes the directional correction (Tingley 2003) - ignores negative depths.
\item  Includes a set of utilities: detection limits estimator (see \S \ref{DetectionLimit}), CB transit predictor, non-linear optimization.
\item  Has better visualization of results for analysis, and has improved robustness, better speed optimization, more 
input control, bug fixes etc.
\item  Allows one to output all the computed CB-BLS values (and not just the maximal one for each orbital period) to properly account for the sample of ?tted models [Ofir 2009].
\end{itemize}

We stress that CB-BLS is a \textit{detection} algorithm, and as such - concisely avoids many complicating elements that are known to affect the shape of the transit light curve, including: stellar limb darkening, motion of EB components along the line of sight (the very basis of RV and eclipse timing techniques), planetary orbital eccentricity and obliquity, non-Keplerian (especially Newtonian) motions, etc. The reason none of these is accounted for is discussed in the last paragraph of the following section.

% ========================
\subsection{CB-BLS with a single depth}
\label{OneDepth}
% ========================

If the EB surface brightness ratio $J$ is known (and $J$ is a natural output of most, if not all, EB modeling tools) then the depth ratio between transits in front of the primary and the secondary EB components will also be $J$, and eq. 1 in paper I can then be re-written with just a single fitted depth as (note the last term):

\begin{equation}
D=\sum_{k}{w_k (x_k - H)^2} + \sum_{i}{w_i (x_i - L)^2} + \sum_{j}{w_j (x_j - JL)^2}
\end{equation}

Here we approximate $L_2$ from paper I to $JL$, which is an approximation to the correct value of $J(L+H)$. The approximation is good as long as $|H| \ll |L|$, which is the case for low duty-cycle signals such as transits. As in BLS (Kov{\'a}cs, Zucker and Mazeh  2002, hereafter KZM) and CB-BLS paper I, minimization allows to analytically compute $H$ (unchanged) and $L$ (revised):
\begin{equation}
H=\frac{-(s_1+s_1)}{1-(r_1+r_2)} \quad \quad , \quad \quad L=\frac{s_1+Js_2}{r_1+J^2r_2}
\end{equation}

where the $r_{1,2}$ and $s_{1,2}$ are simple sums of the data and the weights as in paper I. Plugging these to $D$ gives:
\begin{equation}
D=\sum_{n}{w_n x_n^2} - \frac{(s_1+s_2)^2}{1-(r_1+r_2)}-\frac{(s_1+Js_2)^2}{r_1+J^2r_2}
\end{equation}

Similarly to BLS and to paper I, the first term on the right hand side is constant, leaving the rest as \textit{SR} - the new CB-BLS statistic (times $-1$):

\begin{equation}
SR=\frac{(s_1+s_2)^2}{1-(r_1+r_2)}+\frac{(s_1+Js_2)^2}{r_1+J^2r_2}
\end{equation}

The use of $J$, together with the use of Roche geometry, EB inclination and the directional correction, are all attempts to use all available knowledge about the host EB and the expected transit signal shape to reduce the number of free parameters. For this reason CB-BLS uses the simplest possible box-like transit model and concisely avoids even slightly more complicated models that requires an additional free parameter (such as the trapezoidal transit model, limb darkening, or the other effects mentioned at the end of the previous section). Note however, that the EB model itself can be as complicated as needed - as long as it stays a pre-processing step to CB-BLS it does not add free parameters to the CB-BLS fitting process.

\subsection{Estimating CB-BLS detection limits}
\label{DetectionLimit}

\begin{figure}
\includegraphics[width=0.5\textwidth]{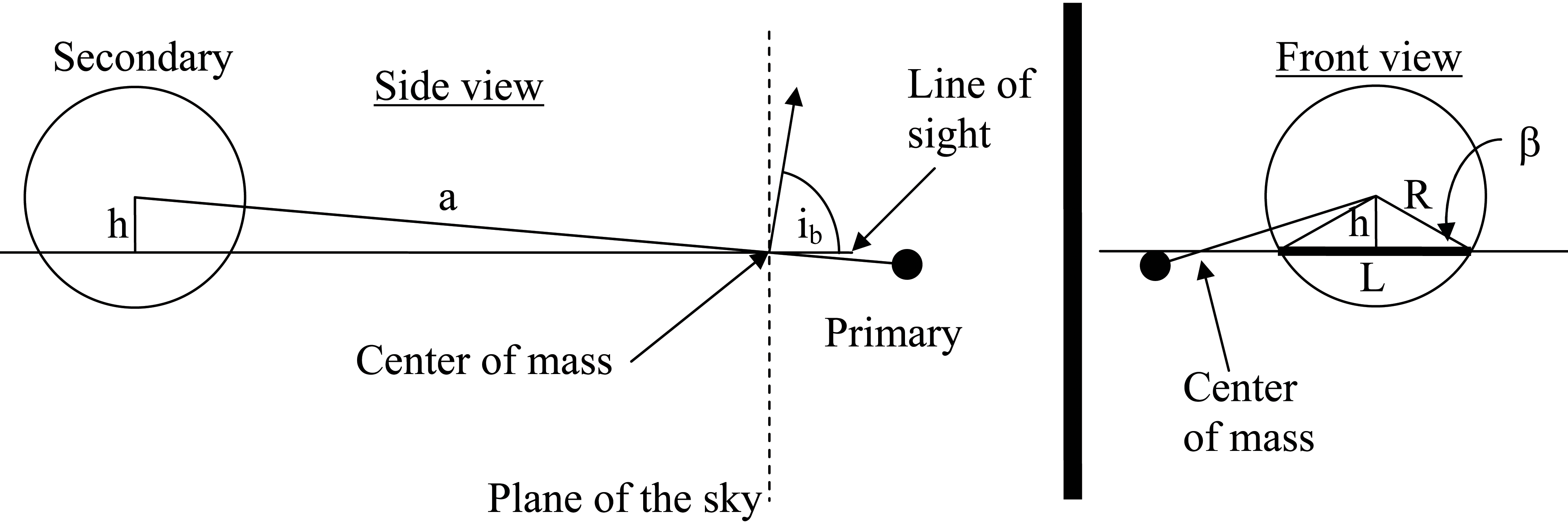}
\caption{Geometrical layout of one component (marked "Secondary") of an EB for the estimation of CB-BLS detection limits The primary is shown only for easyer understanding of the layout. Symbols used are $i_b$ for the binary orbital inclination, $a$ for the radial projection of the component's distance from the center-of-mass, $h$ for the maximum projected distance of the component above the line sight (or more exactly: above the plane containing both the line of sight and the line of nodes), and the angle $\beta$ between a stellar radius and the chord $L$ across the stellar disk cut by the above plane (boldfaced).}
\label{Views}
\end{figure}

Giving robust estimates for detection limits of CB planets is difficult. A full analysis needs to account for all the effects known from planets around single stars, and for a number of additional significant sources of noise, for example: EB modeling errors, which are system-dependant and introduce systematic errors of unknown magnitude. We believe that such a full analysis can only be done with extensive simulations on a system-by-system basis. Still, as long as the approximations assumed in Paper I still hold, especially that the CB planet moves in a circular and edge-on orbit and that the noise is purely Gaussian, one can formulate an analytic detection limit. Fig. \ref{Views} shows the side- and front- (or observers-) views of the geometry of one component of an EB, explaining the different symbols used below.

We begin by considering the effective transit Signal/Noise ratio for single stars defined by KZM:  $\alpha_{single}=\frac{\delta}{\sigma}\sqrt{Nd}$ where $\delta$ is the transit depth, $\sigma$ is the dispersion of the LC, $N$ is the number of data points and $d$ is the transit duty cycle -- the fraction of in-transit data points. If the surface brightness ratio $J$ is known then one can extended $\alpha$ to transiting CB planets; we remind the reader that after regularization the transit depths are only two discrete values with a ratio of $J$. We therefore rearrange to get $\delta\sqrt{Nd}=\alpha \sigma$. In order to account for the fact that two stars can be transited, we change the left side to $\delta \sqrt{N(d_1+d_2)}$ where $d_1, d_2$ are the individual on-transit duty cycles of the components, and in order to account for the different depths of the two components, we make the following changes: $\sqrt{N(d_1\delta_1^2+d_2\delta_2^2)}$, to give a final definition of

\begin{equation}
\alpha_{CB}=\frac{\delta}{\sigma}\sqrt{N(d_1+d_2 J^2)}
\label{AlphaEq}
\end{equation}

where $\delta$ is the regularized depth of the primary transit. When investigating detection limits one must first define what "detection" is. Using the general guidance of Fig. 6 on KZM for $\alpha_{single}$, corrobated by our experience from the Blind Tests of \S \ref{BlindTests}, (where e.g. 'Test3' with $\alpha=11$ was near the limit of detectability, see Fig. \ref{CoToTLimits}), we estimate that robust detections usually have $\alpha \geq 10-12$. We then use the $\alpha_{CB}$ definition to get
\begin{equation}
\delta=\frac{\sigma \alpha_{CB}}{\sqrt{N(d_1+d_2 J^2)}}
\label{DeltaDef}
\end{equation}
We now need to find $d_1, d_2$. For a particular system the duty cycles depends significantly on the exact relation between the orbits of the two binary components and the CB planet -- a small change in the ephemeris can lead to the drastically longer and/or shorter transits. We therefore wish to avoid this problem and consider only the average detection power at a particular planetary orbital period. In analogy to the transit duration calculation around single stars, the average duty cycle for each of the two components is $d=\frac{L}{2 \pi a_p}$ where $L$ is the transit chord lenght of a planet whose orbital plane is spanned by the line of sight (e.g. $i_p$ = $90 ^\circ$) and the line of the component's nodes, as shown in Fig. \ref{Views}.

Since in Fig. \ref{Views} the height $h$ can be found from both the side view as:  $\cos (i_b)=\frac{h}{a}$ and from the front view as: $\sin(\beta)=\frac{h}{R}$ one gets the relation:
\begin{equation}
L=2R \cos(\beta)=2R \sqrt{1-\left({\frac{a \cos(i_b)}{R}}\right)^2}
\label{Leq}
\end{equation}

where $a$ is the radial projection of the component's distance from the center-of-mass (for the definition of $a$, $\beta$, $h$ see Fig. \ref{Views}). Note that $L$ can vanish completely when $a \cos(i_b)>R$ (i.e., the planet passes above/below the binary component without transiting it). We note that $a$ should scale differently for the primary and secondary components according to their mass ratio $q$, and that $a$ is itself a function of binary phase: each component will have the maximum chord length of $2R$ at the orbital nodes (where $a=0$), and lower values -- as low as $L$ of Eq. \ref{Leq} -- at other times. To better estimate $L$ the time-average $\bar a$ for each component should be determined, which for circular orbits (and $L>0$ at all times) is $\bar a=\frac{2}{\pi}a_b$. Using Kepler's third law $\frac{a_p}{a_b}=\left(\frac{P_p}{P_b}\right)^{2/3}$ and substituting back one finds that:
\begin{equation}
d=\frac{R}{\pi a_b}\left(\frac{P_b}{P_p}\right)^{2/3}\sqrt{1-\left({\frac{\bar a \cos(i_b)}{R}}\right)^2}
\label{Deq}
\end{equation}
%For each of the binary components, with only $R$ and $\bar a$ differing for each component, Eq. \ref{Deq} is to be substituted back into Eq. \ref{DeltaDef}.
For each of the binary components the $R$ and $\bar a$ will be different, giving $d_{1,2}$ for the two components -- to be substituted back into Eq. \ref{DeltaDef}.

Since  $i,J,P_b,N,\sigma$ and the ratios: $R/a_b$ and $R/\bar a$ are all given by the EB model and the LC, we now have a detection limit $\delta$ as a function of only our empirical choice of $\alpha_{CB}$ and the EB geometry model -- depending only on the planetary period $P_p$. We note that the derived detection limit is for the best-detectable configuraton of planetary orbits, which are those that are coplanar with the line of sight and at the same position angle as the EB's orbit. Finally, we remark that fulfilling the Gaussian noise requirement may be achieved by quatifiying the correlated noise (e.g. as $\sigma_r$ in Pont \etal 2006) and adding it to the $\sigma$ used in the formulation above.

% ========================
\section{Blind tests to CB-BLS}
\label{BlindTests}
% ========================

CB-BLS was subsequently blind-tested using a detached EB observed during the initial run of CoRoT (CoRoT ID 102806577). Note that the data were not detrended in any way. Notably, CoRoT's orbital period was not removed from the data. Therefore, the tests are done in the presence of strong systematic noises. The "examiner" (HD) created the test data in four steps:
(1) Removal of a constant slope apparent in the data and conversion to relative flux units $(F-F_0)/ F_0$, where $F_0$ is an average of the out-of eclipse flux. The test LCs contain 7213 data points.
(2) Generation of a model light curve (LC) of the EB and removal of
this model from the data.
(3) On the residuals, suppression of the noise\footnote{This step disables 'detections' based on simple subtractions between the various test light curves, all of which had been generated from the same original. For test-sets based on different originals, this step would not be necessary} in a band-pass of periods between 5 h and 12 h. Random noise of similar amplitude was then inserted in that band-pass. This band-pass corresponds to periods a few times longer than the typical duration of transits events; their detectability should therefore not be affected. Also, periodograms of the lightcurves before and after this operation were compared in order to assure that the overall noise characteristics did not change significantly.
(4) Adding to the previous data a model LC that consists of the EB model of step (2) together with a simulated circumbinary planet, using the UTM transit simulator [Deeg 2009]. Steps (3) and (4) were repeated six times, simulating different planets in step (4).

\begin{figure*}
%\centering
\includegraphics[width=0.5\textwidth]{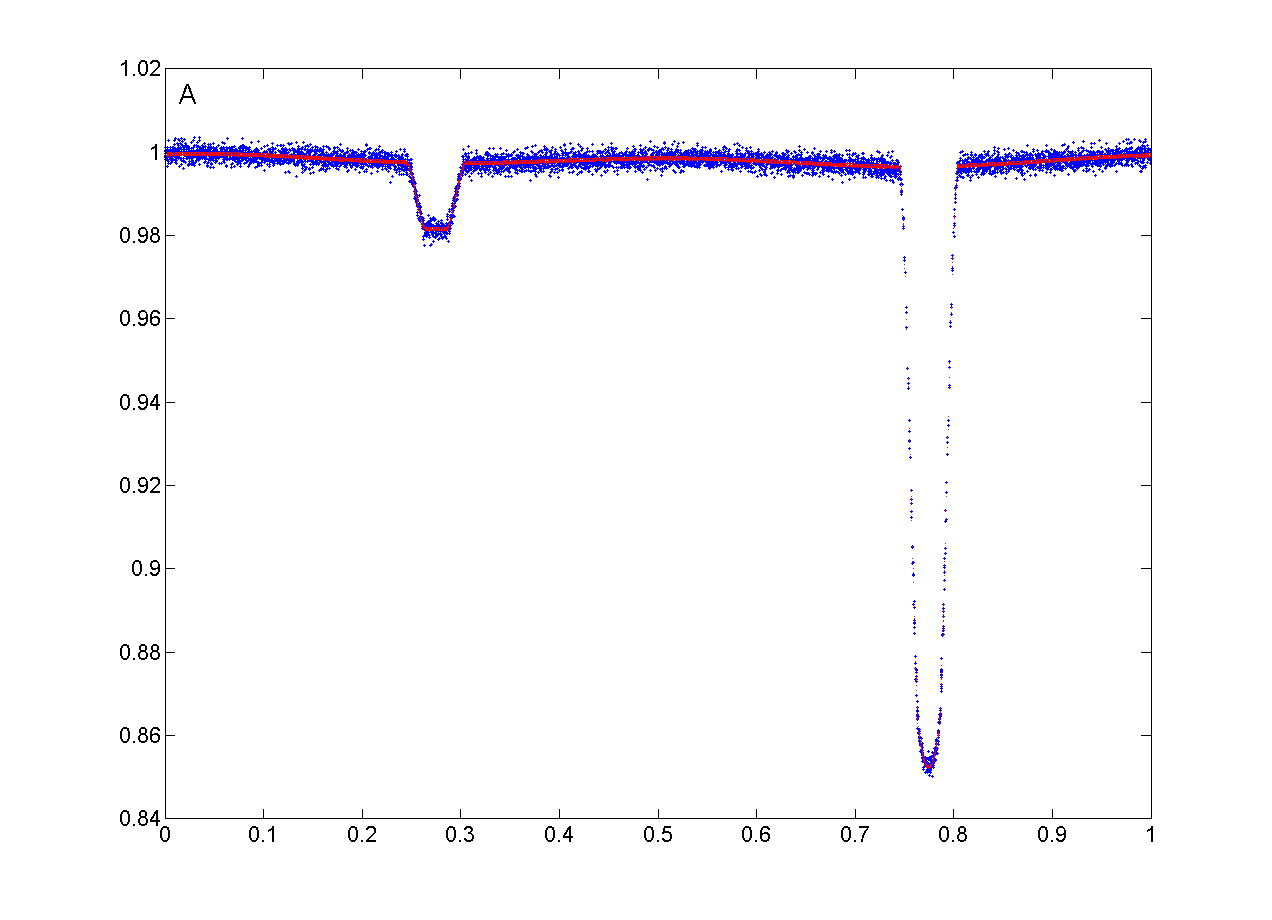}
\includegraphics[width=0.5\textwidth]{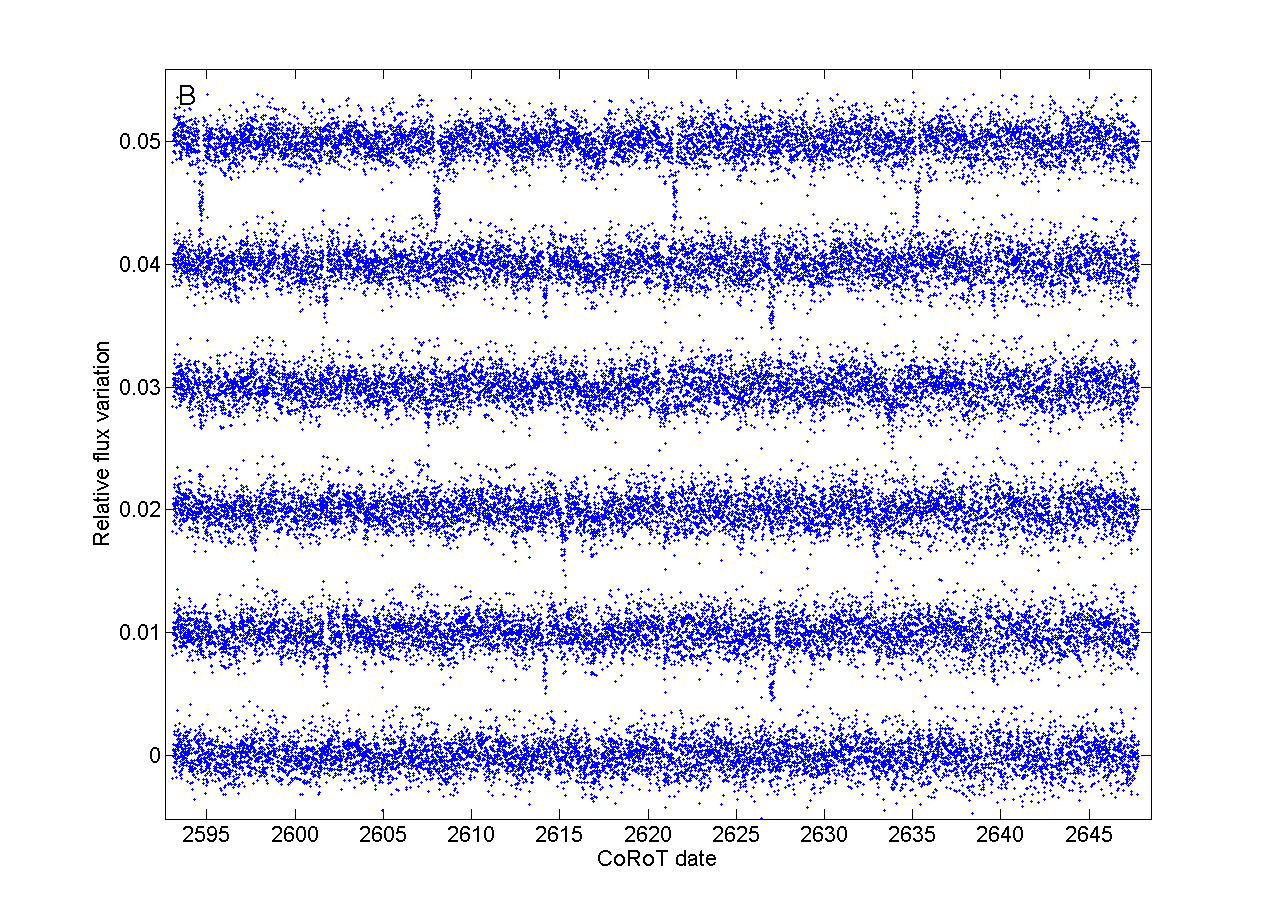}
\caption{Panel A: LC of the EB used in the blind tests (dots) and its JKTEBOP model (line). This LC is the Test5 LC which has no planet added to it. Panel B: the residuals of the six test LCs from Test0 (top) to Test5 (bottom). Each successive residuals LC was shifted by 1\% to aid visibility.}
\label{EB_LC}
\end{figure*}

The CB-BLS "solver" (AO) then received the six test LCs designated Test0 through Test5, and parameters of the physical model of the system. The solver then independently solved the LC using JKTEBOP [Southworth \etal 2004a,b] and applied CB-BLS. For the CB-BLS analysis, only the mass ratio was used out of the full physical model. As written in paper I, if the mass ratio is not known -- it can simply be searched-on as a free parameter. Figure \ref{EB_LC} depicts the EB itself with the JKTEBOP model (from the Test5 LC), and also the six residuals LCs. Note that the residuals have some structure, meaning, they have some correlated noise -- as expected in real data. In Table \ref{EBsystem} we list some of the EB's and simulated planets' physical model parameters that most affect transit detection. Note that the RMS of the residuals is about 1.3mmag. We comment that each of these CB-BLS analyses take approximately 2min on a 3GHz PC (using the natural parameter resolutions presented in paper I), so processing time is neglible, given the small number of targets relative to constant stars.

\begin{figure*}
%\centering
\includegraphics[width=1.0\textwidth]{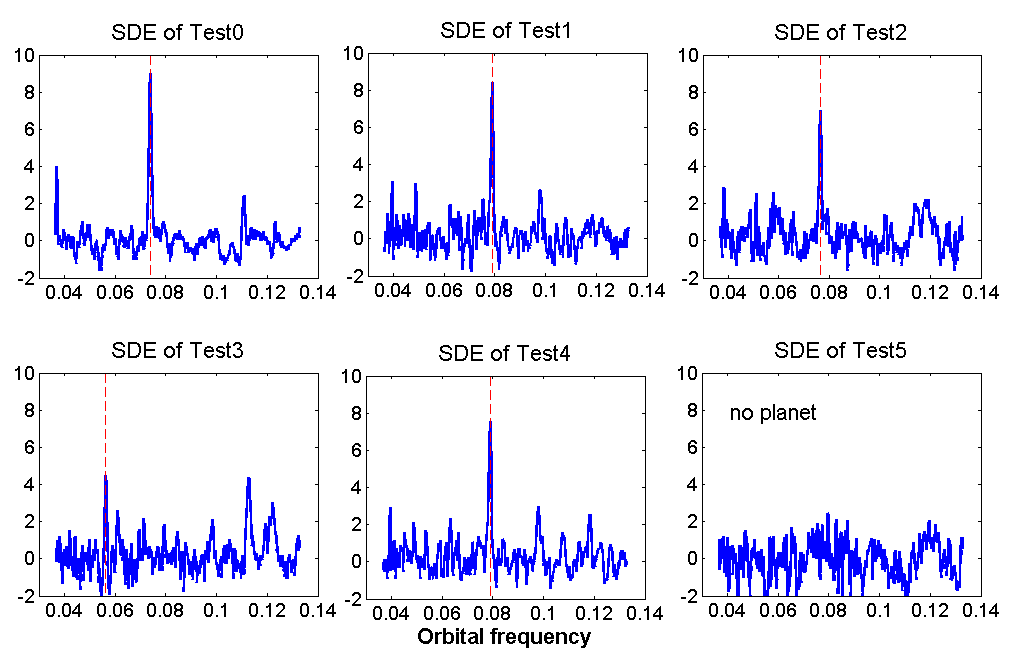}
\caption{The straightened periodograms of the blind tests --- to be read together with the input and output values specified on Table \ref{ResultsTable}.
%Blue line is CB-BLS, and for comparison -- red line is regular BLS on the EB residuals. Each insert shows the region around the correct frequency, except Test5 which does not include a planet.
The correct frequency is marked with a dashed vertical line. The highest frequency ($f\sim 0.136 d^{-1}$) corresponds to a period just twice that of the binary -- near the instability limit. The periodograms are normalized to the their own RMS (as described in Paper I) -  thus, a value of 5 is a $5\sigma$ detection of a certain period.}
\label{SDEgallary}
\end{figure*}

\begin{table}
\caption{Main EB- and CB planet- parameters that affect the planet's transit.}
\begin{tabular}{l r l}
\hline
Parameter 		& Value 				& Unit \\
\hline
Period			& 3.6670288245 & [d]\\
Inclination		& 86.6699982 & [deg]\\
$R_1$				& 1.856 	& [$R_\odot$]\\
$R_2$				& 0.6838 & [$R_\odot$]\\
$L_1$				& 14.620 & [$L_\odot$]\\
$L_2$				& 0.2500 & [$L_\odot$]\\
RMS of residuals & $\sim 1.3$ & 	[mmag] \\
\hline
Test0 $R_p$			&1.09	&	[$R_J$]\\
Test1 $R_p$			&0.95	&	[$R_J$]\\
Test2 $R_p$ 		&0.78	&	[$R_J$]\\
Test3 $R_p$ 		&0.79	&	[$R_J$]\\
Test4 $R_p$ 		&0.95	&	[$R_J$]\\
Test5 $R_p$ 		&none	&	[$R_J$]\\
\hline
\end{tabular}
\label{EBsystem}
\end{table}

The six tests signals were such that Test0 was relatively easy and not blind (it was used primarily to make sure the test procedures are working) - but still the simulated system was a physically-possible system, while Tests 1-5 were completely blind. Fig. \ref{SDEgallary} shows the resultant CB-BLS periodograms
% (and for comparison - BLS periodograms too)
 and Table \ref{ResultsTable} gives the resultant best CB-BLS fits values. All six test LCs were correctly identified, where notably Test5 was correctly identified only when the single-depth SR of \S \ref{OneDepth} was used.
% The BLS peaks of Tests 1-4 were off the correct frequency, significantly wider (i.e., less precise) and with generally lower significance than the CB-BLS peaks from the same data.
 Note that the simulated planets had their the inclination and position angle varied to some extent and that their LCs also account for limb darkening - but since these none of these features are derivable from CB-BLS they are not given in this comparison. 

\begin{figure}
\includegraphics[width=0.5\textwidth]{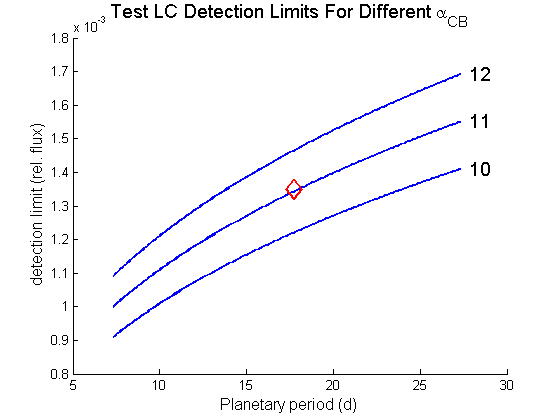}
\caption{Detection limits for $\alpha=10,11,12$ of the CoRoT test LC (sold lines) and the Test3 signal as it was detected (from Table \ref{ResultsTable}).}
\label{CoToTLimits}
\end{figure}

\begin{figure*}
%\centering
\includegraphics[width=1\textwidth]{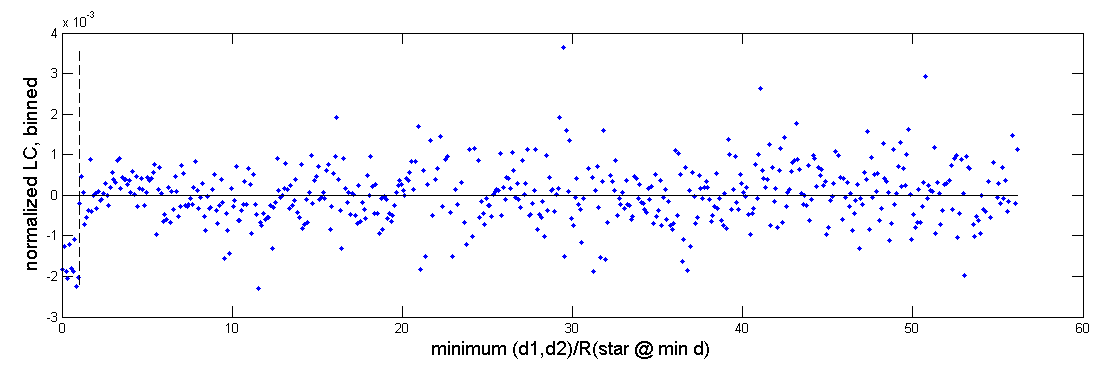}
\caption{The Test3 LC residuals folded against scaled-distance for the best-fit model (see text for details). Readers who are familiar with phase-folded LCs can view of this figure as yet another fold of the LC about the center of the transit, so that both ingress and egress fall on scaled-distance $\sim 1$, and all in-transit points are expected below scaled-distance of 1 (to the left of the vertical dashed line). To aid visibility the data is binned to 0.1 units of this scaled-distance. Despite the fact that Test3 was the shallowest signal in this blind tests series CB-BLS clearly separates in-transit from out-of-transit points.}
\label{Test3}
\end{figure*}

While simple phase-folding the LC will not separate well in- and out- of transit points (which is the motivation behind CB-BLS), a "folded" graphical representation of the CB planet's transits is possible (and instrumental for signal analysis): in-transit points are expected when the sky-projected distance between the planet and one of the components is closer than that component's radius. For that reason the data can be folded against the minimum distance of the planet from each of the EB components - and that distance is scaled by that (i.e., the closer) star's radius. The result is that all in-transit points are expected below scaled-distance of 1, and all out-of-transit points are expected at scaled-distances grater than 1. Fig. \ref{Test3} shows the LC of Test3 folded in this way, which can also serve as an illustration for the ability of CB-BLS to detect even shallow transits significantly. Note that this simulated planet caused only three primary transit events, where a primary transit is only $\sim 1.4$ mmag deep (to be compared with RMS dispersion of 1.3 mmag). This depth corresponds to a planetary radius of $0.73 R_J$ in an EB with a primary component with a radius of $1.856 R_\odot$. Scaling this depth to $1 R_\odot$ (and ignoring the much-fainter secondary component) gives $0.39R_J$ - smaller than all known transiting planets but the unique super-Earth CoRoT-7b [Rouan \etal,
2009]. As seen on Fig. \ref{SDEgallary} the Test3 LC is significantly detected, but is not far from the limit of reliable detection. Indeed, one can also apply the detection limits procedure (\S \ref{DetectionLimit}) to the CoRoT test LC too - and Fig. \ref{CoToTLimits} shows these for various $\alpha_{CB}$ values with the detected signal for reference -- and the detected signal is indeed just above or just below the detection limit, depending on the choice of $\alpha_{CB}$. With the successful detection on all test signals and the agreement between the theoretical detection limits and actual detection, we conclude that CB-BLS passed the blind tests successfully and is ready for use on real data.

\begin{table*}
\caption{The input values of the simulated CB planets ("Examiner" rows) and their respective solution ("Solver" rows). The possible detection levels are: "+++" (Strong and secure), "++" (Probable. Noisy), "+" (Weak detection) and "no planet". The depth of the secondary transit is omitted since it is simply $J$ times the primary transit.}
\begin{tabular}{|p{1.0cm}|l|c|p{1cm}|p{1.8cm}|p{1.5cm}|p{1.4cm}|p{0.8cm}|p{4.3cm}|}
\hline

Test dataset &	Attribute 		&	Detection&	Period \ [d]& Baricenter crossing \ [CoRoT JD]	& Depth (primary) [mmag]	&	In-transit points &	$\alpha_{single}$	& Comment\\
\hline
\raisebox{-1.7ex}{Test0}	&	Examiner	&				&	13.54			&  2594.54					&	4.683			&	&			&\\
									&	Solver	&	+++		&	13.545		&	2594.525					&  4.768 		&	148 &	44.0		& Not blind.\\
\hline
\raisebox{-1.7ex}{Test1}	&	Examiner	&				&	12.673		& 	2563.567					& 2.749 			&	&		& \\
									&	Solver	&	+++		&	12.667		&	2595.251		 			& 2.494 			&	145 & 22.8		& A bug (since then fixed) caused the detected reference time to be when the planet crossed the baricenter \textit{behind} the EB system, causing a half-integer difference (2.5 periods) from the correct one.\\
\hline
\raisebox{-1.7ex}{Test2}	&	Examiner	&				&	13.08 		& 	2594.54					& 1.853			&	&		& \\
									&	Solver	&	+++		&	13.091		&	2594.513 				& 1.560 			&	248 &	18.6		& \\
\hline
\raisebox{-1.7ex}{Test3}	&	Examiner	&				&	17.748		& 	2597.54 					& 1.901			&	&		& \\
									&	Solver	&	+++		&	17.738		&	2597.571					& 1.349  		&	132&	11.7		& \\
\hline
\raisebox{-1.7ex}{Test4}	&	Examiner	&				&	12.673		& 	2563.567					& 2.749			&	&		& \\
									&	Solver	&	+++		&	12.652		&	2595.295				& 2.579 				&	143 &	23.4		& See comment to Test1\\
\hline
\raisebox{-1.7ex}{Test5}	&	Examiner		&	no planet&		-				& 		-						&	-		&	&			&\\
									&	Solver	&	no planet&		-				&		-						&  -			&	0 & 0	& A very weak signal was initially suspected before using the single-depth analysis. That periodogram peak completely disappeared when this feature was added and used.\\
\hline
\end{tabular}
\label{ResultsTable}
\end{table*}

%First line: my units (binary semimajor axis). second line: converted to
%Rsun.
%      |  Test0  |  Test1  |  Test2  |  Test3  |  Test4  |  Test5   |
%      ==============================================================
%R_p   | 0.00924 | 0.00664 | 0.00548 | 0.00539 | 0.00677 | 0.003402 |
%      | 0.126   | 0.0903  | 0.0746  | 0.0733  | 0.0920  | 0.046282 |
%true: | 0.128   |                                       | formal result.
%                                                        | again, I see
%                                                        | nothing there

% ========================
\section{Application of CB-BLS to real data}
% ========================

% ========================
\subsection{TrES Lyr1 field}
% ========================

The photometry of the TrES survey Lyr1 field [O'Donovan \etal 2006, Dunham \etal 2004] is freely available at the NASA NStED website \footnote{http://nsted.ipac.caltech.edu/}, but this photometry is after TrES's detrending, and so can't be used as-is for our purpose. We therefore asked the above authors for the raw photometry of the same data, and we process below only the raw Sleuth data - which contributes $\sim 11,000$ data points of the total $\sim 15,000$ of the TrES Lyr1 dataset.

\begin{figure}
\includegraphics[width=0.5\textwidth]{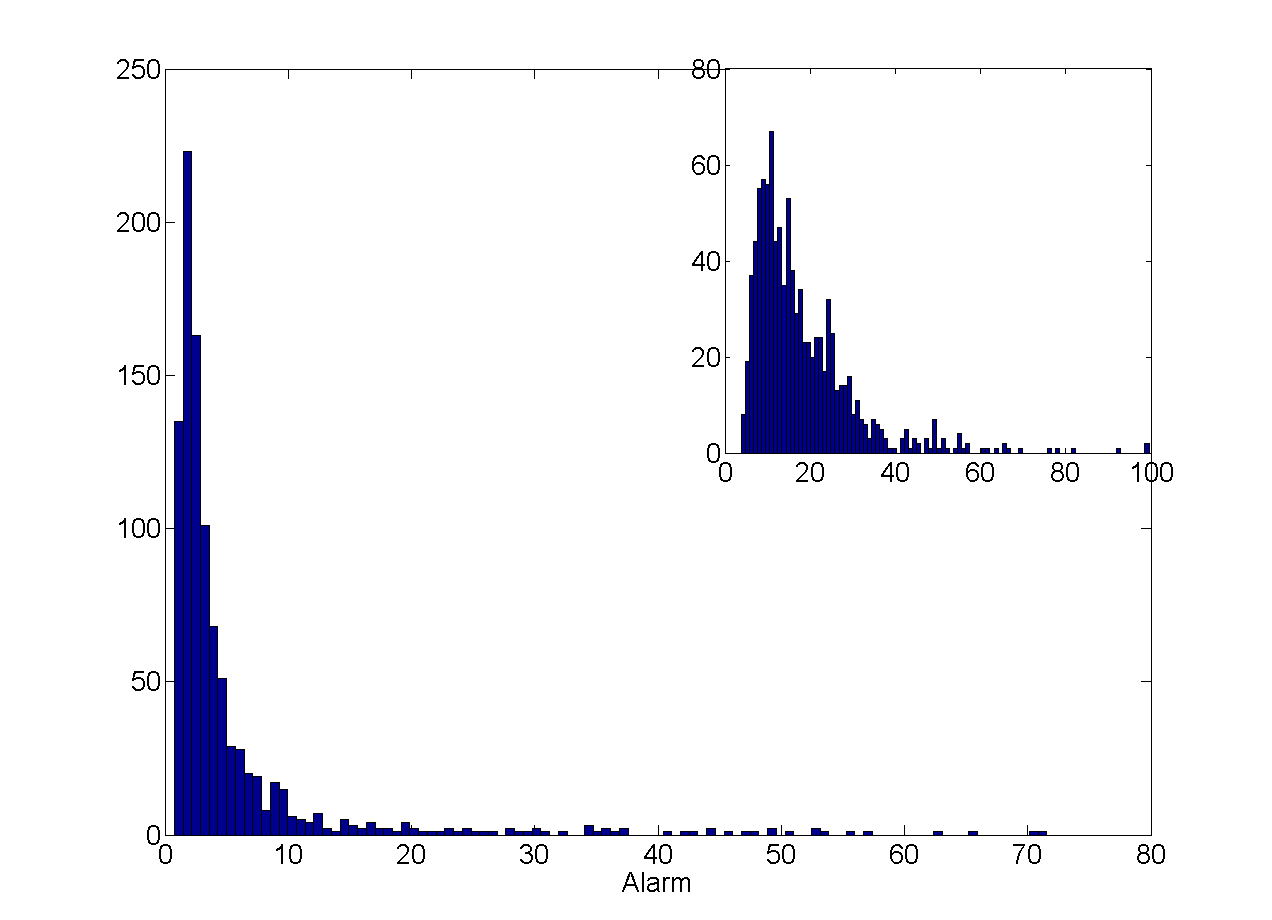}
\caption{The Alarm statistic for 1000 objects (all objects in some magnitude range) of the TrES Lyr1 dataset. The insert shows the lower part of the Alarm distribution before the first iteration for variable-stars filtering (see text). After we used only stars with Alarm$<35$ to calculate one SYSREM effect - the Alarm distribution of the same objects looked as shown on the main panel.}
\label{Alarm}
\end{figure}

Here we present our first application of CB-BLS to photometry from a real wide-field survey  and the following pre-processing scheme is the template for our future applications of CB-BLS. The main difference between the scheme below and the corresponding pre-processing steps of other transit searches is the ability to not only detrend intrinsically-constant stars but also to detrend variable stars by iteratively detrending their residuals. This approach was first proposed by Kov{\'a}cs \etal (2005) for their reconstructive TFA algorithm of transit signals, but here we aim to de-trend EBs, which show large variety of variations, and so - the bin-averaging method proposed by Kov{\'a}cs \etal (2005) is ill-suited for the current problem. The pre-processing has four main stages:

\textbf{1. Determination of the systematic effects}: This is very similar to what is done routinely on transit surveys. Firstly, outlier data points are removed form all light curves by sigma clipping each light curve around a small-window (5 points) median filter. Next, we wish to determine a few SYSREM effects [Tamuz \etal 2005] - but we must make sure that variable stars are not part of the set of stars that is used to determine the effects. To that end we use the Alarm statistic [Tamuz \etal 2006] as a general variability statistic (used similarly to the Stetson J statistic [Stetson 1996]) in the following manner (Note that in general Alarm will have a smaller value for less-systematic input): 1) We compute Alarm for all stars 2) we determine the sub-set of constant stars using an Alarm maximal value that corresponds to the bulk of the stars (i.e., many stars that change similarly, so they probably don't have intrinsic variability). See Fig. \ref{Alarm} for an example. 3) The SYSREM effect(s) are computed using only this subset of constant stars (one effect in the above example) and the correction is applied to all the stars, 4) Alarm is re-computed. Since now the bulk Alarm distribution is narrower (e.g., main panel of Fig \ref{Alarm}) it allows to better filter true variables - so using a lower Alarm maximal value the set of constant stars is re-determined. Steps 2-4 can be repeated several times till the set of constant stars converges. Note that in each iteration the effects must be always computed from the same raw data and not from the SYSREM-corrected data. This procedure also improves the signal-to-noise ratio of the SYSREM effects. For the TrES Lyr1 data we used the final set of constant stars to calculate the three SYSREM effects, with ultimate precision similar to that of the TrES survey.

\begin{figure}
\includegraphics[width=0.5\textwidth]{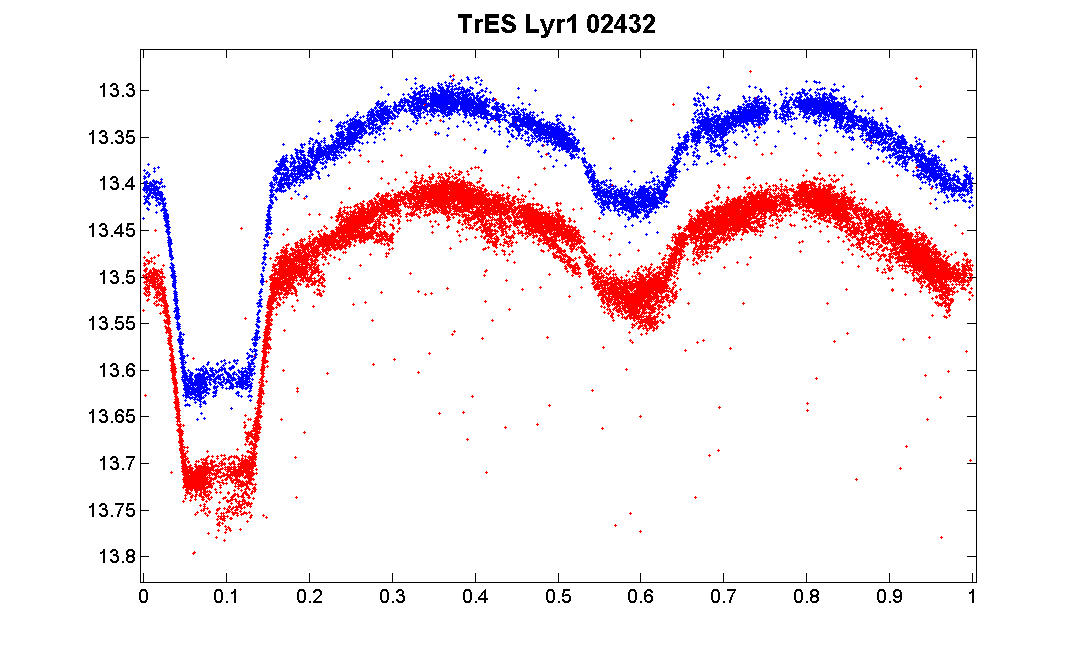}
\caption{Two light curves of star 02432 from the TrES Lyr1 field. Bottom, Red LC: the light curve as posted on the public NStED database: it was detrended with no special care for intrinsic variability. Top, Blue LC: the same raw data processed as described in the text - with cleaning procedures applied only to the residuals. Note that transit detection in LCs such as the red LC is nearly impossible.}
\label{02432}
\end{figure}

\textbf{2. Detrending variable stars}: The constant stars allowed us to determine the systematic effects well - and from now on the effects are assumed known and will not be changed. To correct variable stars for these systematic effects we will have to apply the same correction determined above to the residuals around the smoothed LC of the target variable star by iteratively smoothing the target LC, detrending the residuals, adding back the corrected residuals to the previous smooth and then re-smoothing the LC, till convergence While this procedure is sensitive to the exact smoothing technique, when the smoothing parameters fit well for a given LC the results are quite significant for transit detection (see Fig. \ref{02432}), and allow us to reduce the scatter of the LC around the model to the same level as the scatter of LCs of constant stars of similar brightness.

\begin{figure}
\includegraphics[width=0.5\textwidth]{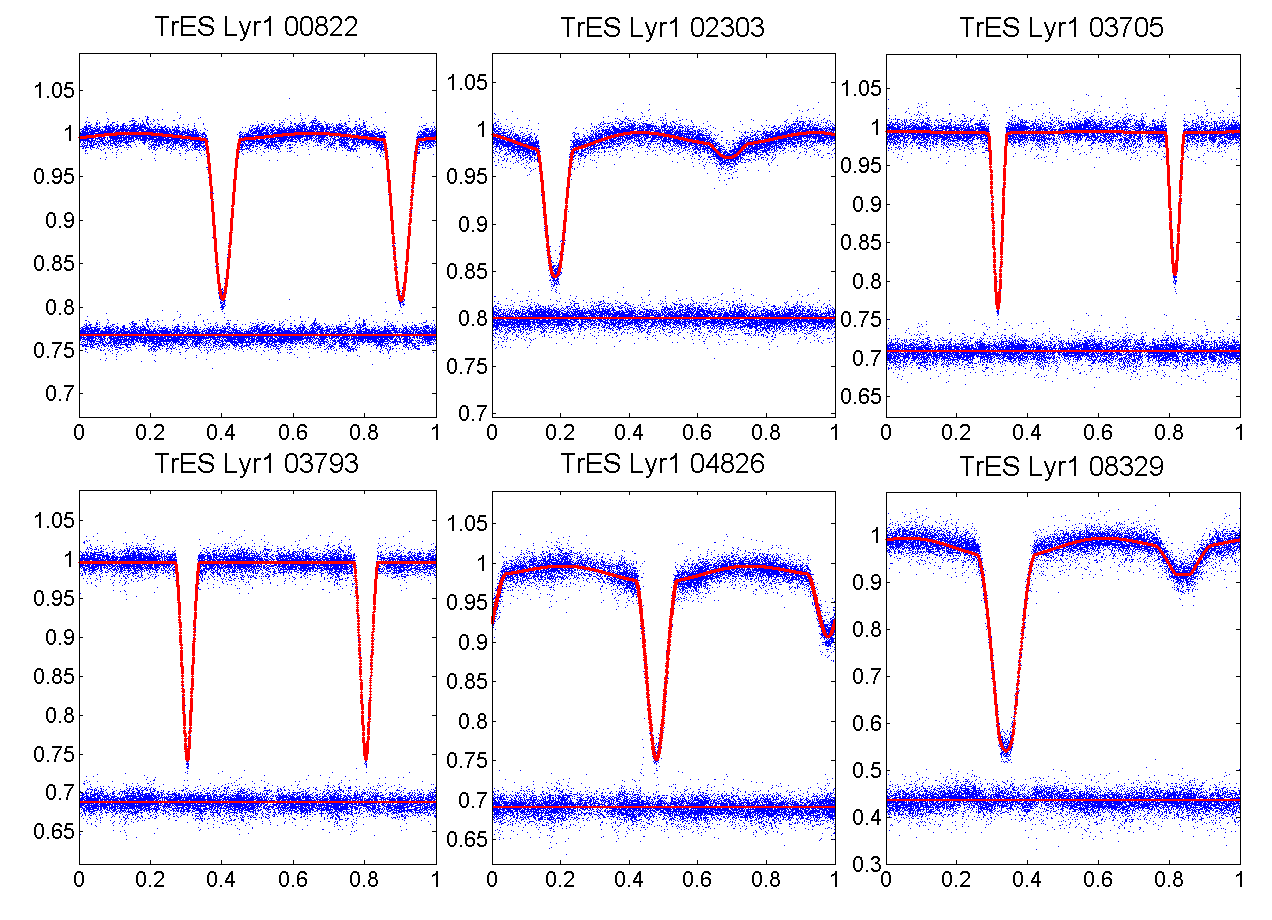}
\caption{More examples of the EBs from the TrES Lyr1 field. These EB LCs were generated using the process discussed in the text for the proper detrending of variable stars. This, together with good modeling, resulted in extremely small systematics in the residuals across the $>11,000$ data points. The object ID at the top of each panel is the NStED designation.}
\label{Galery2}
\end{figure}

\textbf{3. EB solution:} Light curves for eclipsing binaries in the Lyr1 field were analyzed by the NDE model.  Errant observations were picked out by eye and eliminated before analysis.  Accurate dates on minima were determined from the data by using the method of Kwee \& van Woerden (1956).  The resulting periods and zero epochs are listed in Table \ref{EBsParams}, along with the fitted orbital parameters. In most cases, the fitted light curves were relatively insensitive to the assumed values of the limb-darkening coefficients ($x_p$ and $x_s$) and to the mass ratio ($q$), which was guessed based on preliminary values of the light ratio. 5 of the 17 light curves showed significant eccentricity.

\begin{table*}
\caption{Results of fitting light curves of bright EBs from the TrES Lyr1 field. Boldfaced star names (NStED ID) indicate that CB-BLS was subsequently applied. The other columns are: the period $P$ and reference epoch $E_0$ (in HJD-2453000), the surface brightness ration $J_s$, the primary, secondary radii and their ratio $r_p$, $r_s$ and $k$, the orbital eccentricity, inclination and argument of periastron passage $i$, $e$ and $\omega$, the light fraction of third light, primary and secondary components $L_3$, $L_p$ and $L_s$, the linear limb-darkening coefficients $x_p=x_s$, the mass ratio $q=\frac{m_s}{m_p}$, the RMS of the data around the model $\sigma$, and the number of data points used in the fit $N$.}
\tiny
\begin{tabular}{l l l l l l l l l l l l l l l l l}
\hline
Star ID         &$P$ &$E_0$        &$J_s$   &$r_p$    &$r_s$     &$k$     &$i$    &$e$     &$\omega$ &$L_3$    &$L_p$    &$L_s$    &$x_p,x_s$ &$q=\frac{m_s}{m_p}$    &$\sigma$ &$N$ \\
& [d] & & &[$a_b$] &[$a_b$] & &[deg] & &[deg] & & & & & & [mmag]\\
\hline
00442  &1.0620113 &544.9157 &0.577 &0.252 &0.180 &0.714 &85.3 &0.045 &-90.3  &0.042 &0.775 &0.225 &0.60 &0.863 &7.255  &11011 \\					% Lyr1\_01\_359
00822  &2.42759   &562.8633  &1.03  &0.157 &0.190 &1.21  &79.6 &0.004 &-100.0 &0     &0.397 &0.603 &0.60 &1.0   &7.112  &11028 \\					% Lyr1\_01\_687
\textbf{01199} &2.696428  &555.7985  &0.374 &0.174 &0.083 &0.477 &90   &0.003 &-63.0  &0.032 &0.922 &0.078 &0.60 &0.5   &7.804  &10997\\		% Lyr1\_02\_1013
01679 &1.401052  &551.8386  &0.103 &0.292 &0.216 &0.74  &66.5 &0     &0      &0.239 &0.947 &0.053 &0.60 &0.67  &7.288  &11151 \\					% Lyr1\_02\_1439
\textbf{02303} &1.387197  &546.7336 &0.049 &0.280 &0.100 &0.358 &79.6 &0.072 &83.8   &0     &0.994 &0.006 &0.60 &0.368 &7.978  &11074 \\		% Lyr1\_02\_1986
\textbf{03705} &4.703549  &560.7989  &0.793 &0.113 &0.071 &0.63  &85.5 &0     &0      &0     &0.760 &0.240 &0.60 &0.95  &10.146 &11072 \\		% Lyr1\_04\_3204
03708 &0.428355  &552.7032 &0.95  &0.377 &0.354 &0.94  &60.9 &0     &0      &0     &0.55  &0.45  &0.60 &0.95  &14.256 &11087 \\					% Lyr1\_04\_3211
\textbf{03793} &3.472262  &545.7911 &1.005 &0.113 &0.100 &0.88  &87.4 &0     &0      &0.32  &0.56  &0.44  &0.48 &0.95  &9.754  &11148 \\		% Lyr1\_04\_3285
\textbf{04063} &0.876825  &549.8667  &0.000 &0.398 &0.179 &0.45  &69.7 &0     &0      &0.153 &1.000 &0.000 &0.60 &0.29  &11.848 &11096 \\		% Lyr1\_04\_3520
\textbf{04826} &1.25042   &596.8055  &0.302 &0.240 &0.155 &0.65  &80.8 &0     &0      &0.20  &0.89  &0.11  &0.60 &0.85  &10.674 &10410 \\		% Lyr1\_05\_4168
05140 &0.903320  &554.9364  &0.010 &0.339 &0.203 &0.714 &84.8 &0     &0      &0.249 &0.996 &0.004 &0.60 &0.65  &24.129 &10214 \\					% Lyr1\_05\_4431
05911 &0.2932419 &545.7257  &0.66  &0.395 &0.355 &0.90  &58.4 &0     &0      &0.625 &0.656 &0.344 &0.60 &0.95  &14.555 &10679 \\					% Lyr1\_06\_5102
05926 &0.7085494 &545.8493  &0.075 &0.455 &0.329 &0.72  &55.4 &0     &0      &0     &0.964 &0.036 &0.60 &0.80  &14.249 &10597 \\					% Lyr1\_06\_5114
06613 &0.7445117 &546.8838  &0.54  &0.55  &0.44  &0.80  &41.8 &0     &0      &0     &0.75  &0.25  &0.60 &0.80  &16.191 &11095 \\					% Lyr1\_06\_5706
\textbf{06825} &1.801803  &551.9185  &0.133 &0.264 &0.242 &0.917 &88.9 &0.205 &88.6   &0     &0.887 &0.100 &0.40 &0.70  &18.272 &10647 \\		% Lyr1\_06\_5887
07919 &6.5233411 &549.9007  &0.272 &0.260 &0.249 &0.956 &67.3 &0     &0      &0     &0.792 &0.208 &0.60 &0.40  &18.067 &10426 \\					% Lyr1\_07\_6493
\textbf{08329} &1.322684  &593.8688  &0.090 &0.302 &0.186 &0.615 &89.3 &0     &0      &0     &0.967 &0.033 &0.60 &0.60  &17.262 & 9690 \\		% Lyr1\_08\_7179
\hline
\end{tabular}
\label{EBsParams}
\end{table*}

\textbf{4. Generation of final EB LC}: Almost identically to the "Detrending variable stars" step - the raw LC is detrended using the solved EB model. Now the model is final so multiple iterations are not needed.

\begin{figure}
\includegraphics[width=0.5\textwidth]{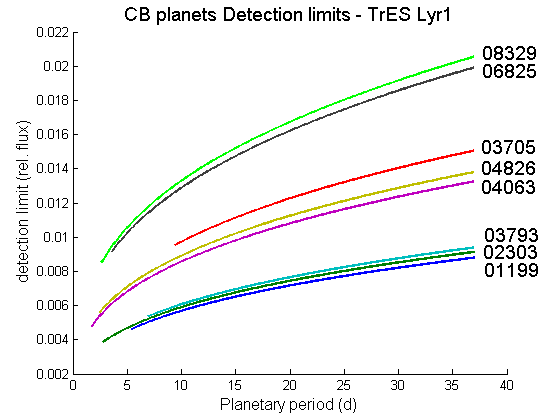}
\caption{The $\alpha_{CB}=12$ detection limits for each of the TrES EBs marked in Table \ref{EBsParams} --- i.e., the smallest depth of primary transits caused by a hypothetical CB planet in the systems that is expected to be reliably detected with CB-BLS. The detection limit was calculated as explained in \S \ref{DetectionLimit}. The planetary periods used are from twice the host EB period to half the survey time span.}
\label{TrESLimits}
\end{figure}

Fig. \ref{Galery2} depicts further examples of EBs from the TrES Lyr1 field, showing that the generation of multiple near-systematics-free EB light curves from wide-field transit surveys using the procedure above is indeed possible. The low level of systematics will help CB-BLS to be less likely to find spurious signals. All in all, about two dozen detached or semi-detached EBs were bright enough (similarly-bright constant stars had ultimate precision of $\sim 2\%$) to be processed as above. Of the two dozen EBs identified 17 were fitted, and for 8 of them planet detection seemed possible (the other EBs exhibit either low inclination angles, evidence of spots or period change), and CB-BLS was applied -- but we did not detect any significant signal. The $\alpha_{CB}=12$ detection limits for each of the objects are given in Fig. \ref{TrESLimits}.

We comment that since there were about 7000 objects in the relevant magnitude range one may expect about 70 EBs to be present in the data, and thus a yield of two dozen EBs may seem too low. However, a similar number of EBs were actually identified but prioritized very low and not processed further (such as EBs with long periods, grazing EBs, EBs with low SNR, EBs with additional signals overlaid, non-eclipsing contact EBs, etc.). Considering that further EB's were missed due to the window function - the total fraction of EBs seems roughly compatible with CoRoT's $\sim 1\%$ above.

% ========================
\subsection{Other targets}
% ========================

While wide-field transit surveys such as CoRoT will probably be the main source of EBs for future CB-BLS analyses (as demonstrated above), there are currently a number of open literature papers on EBs that have a long enough and accurate enough LCs to possibly allow the detection of transiting planets. These papers focus on the EBs and so have the added benefit of having the EBs (usually) fully analyzed - including spectroscopic determination of the mass ratio $q$ (which allows us to remove one free parameter, and shorten the CB-BLS run time). Note that for some of these objects the new multi-band CB-BLS feature is required. Also note that while the blind test were conducted on a $\sim 50d$ long LC, these objects were observed over much longer periods -- typically several years. This necessitates a correspondingly higher frequency resolution, and consequently -- significantly longer CB-BLS run time.

\begin{figure}
\includegraphics[width=0.24\textwidth]{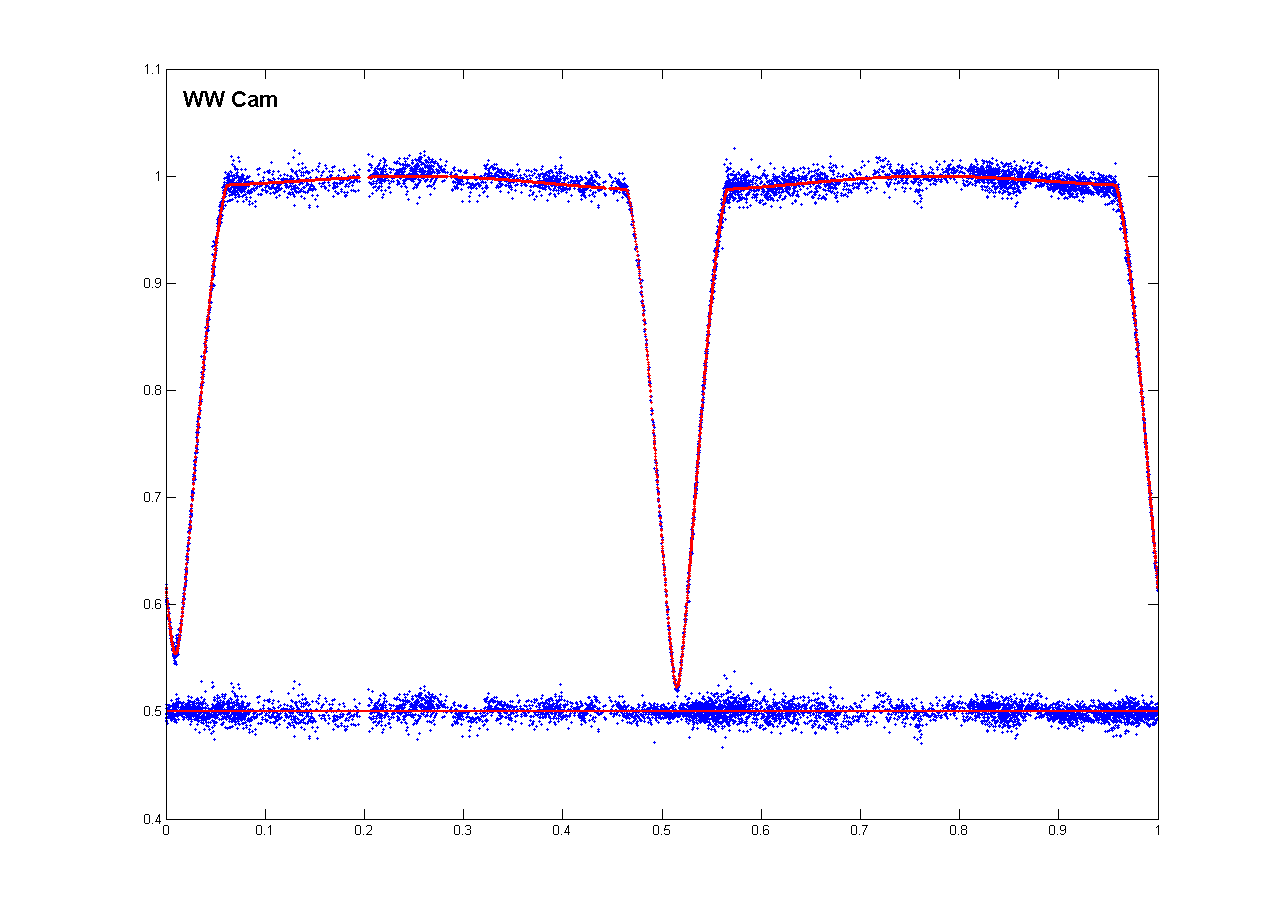}
\includegraphics[width=0.24\textwidth]{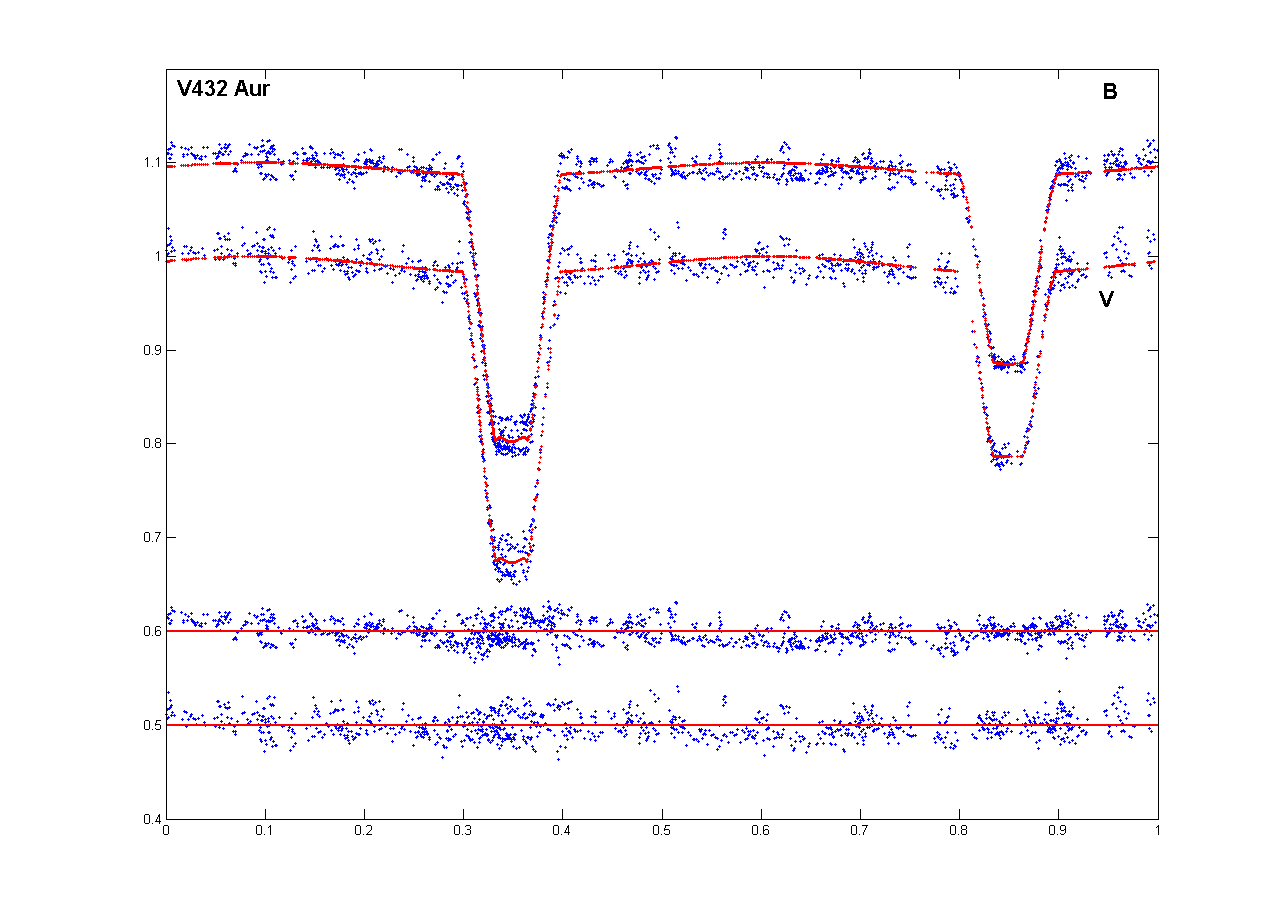}
\includegraphics[width=0.24\textwidth]{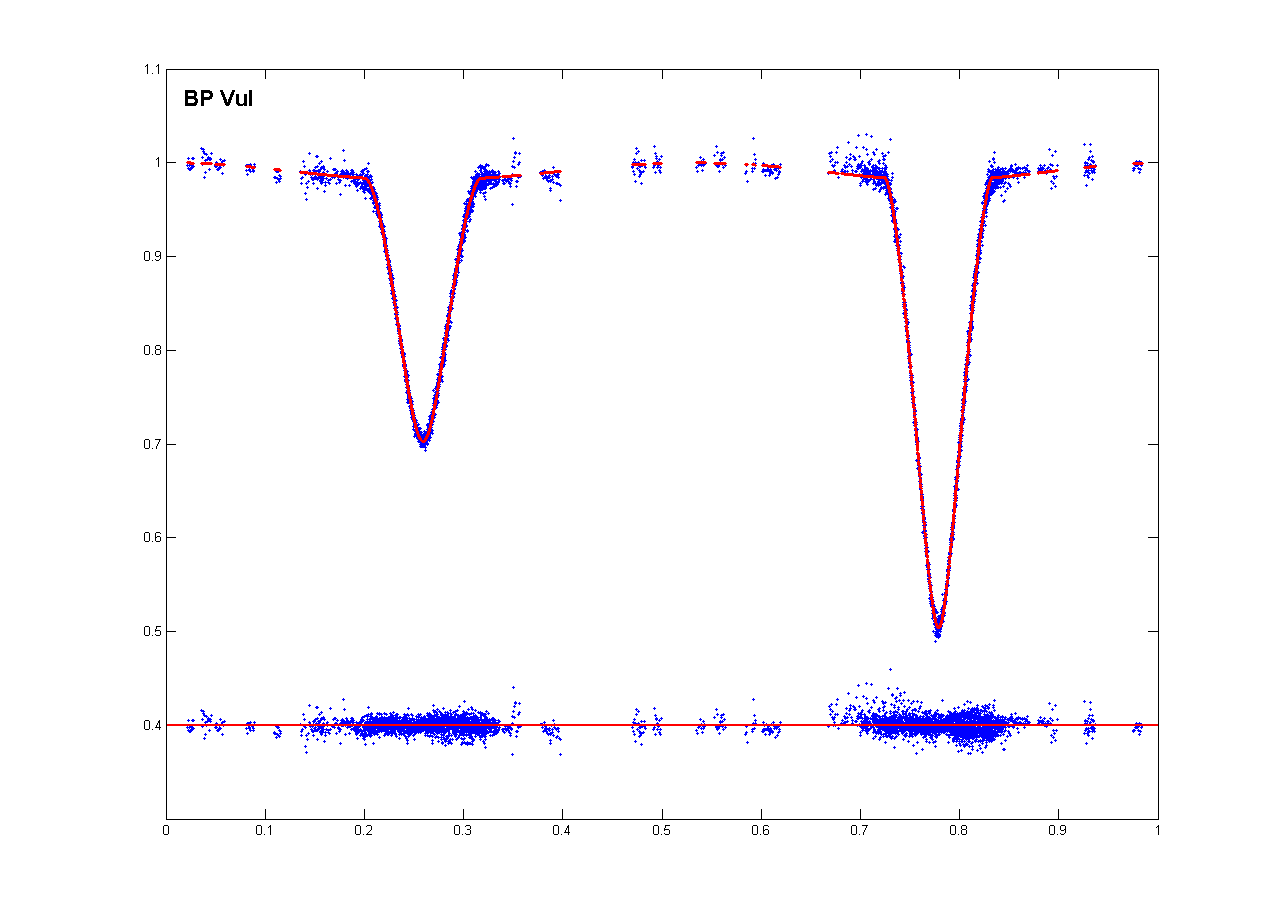}
\includegraphics[width=0.24\textwidth]{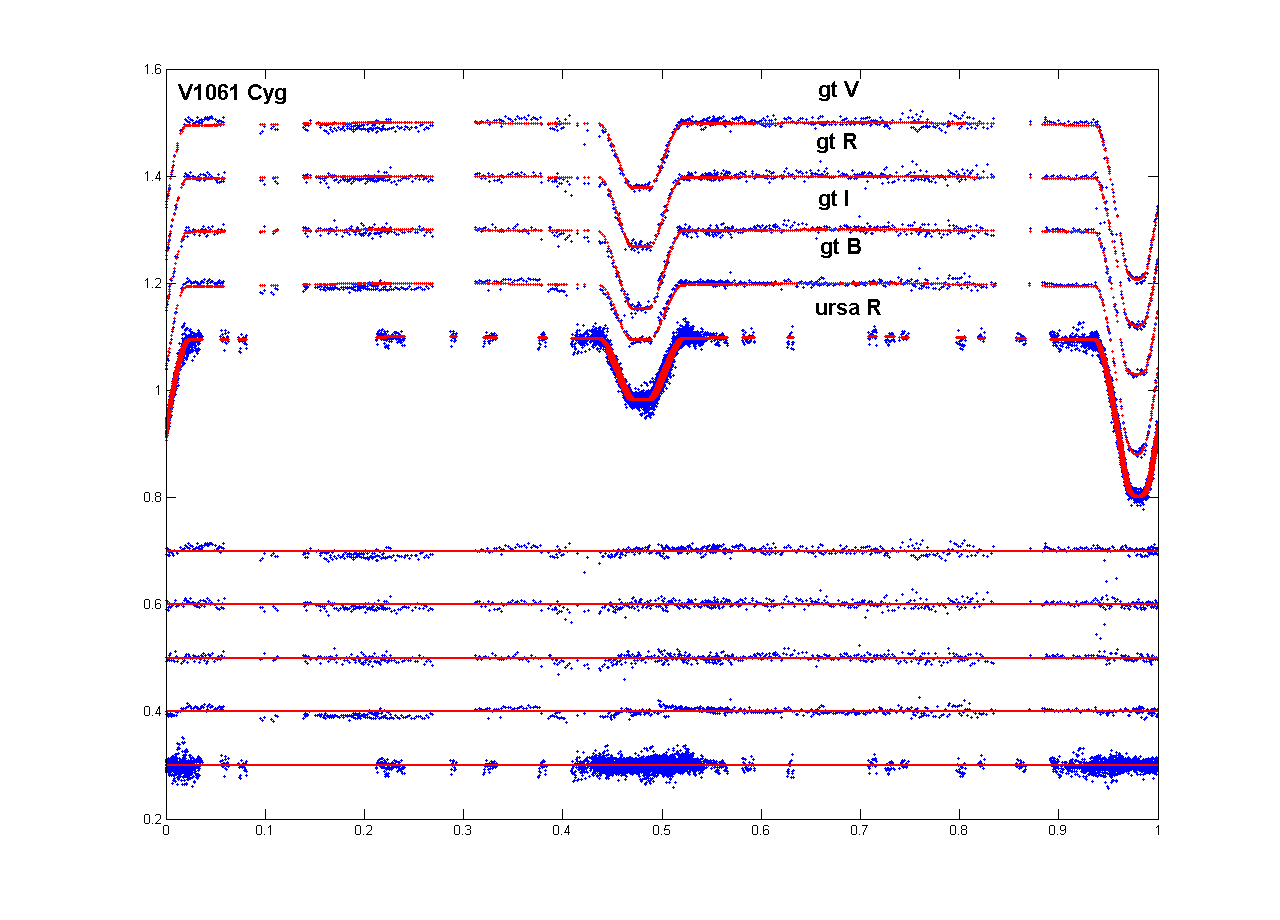}
\caption{Some of the open literature EBs to which the CB-BLS analysis was applied. Each EB light curve is phased to it's orbital period with the best-fit model over plotted and the model residuals drawn below.}
\label{Others}
\end{figure}

Light curves of WW Cam (Lacy \etal 2002) and V1061 Cyg (Torres \etal 2006) were taken from the literature and fitted by the NDE model (Etzel 1981, Popper \& Etzel 1981) as described above. LC parameters for those stars are listed in the referenced articles. Light curves of V432 Aur [Siviero \etal 2004] and BP Vul [Lacy \etal 2003] were also taken from the literature and fitted with JKTEBOP -- giving LC parameters similar to the ones listed in the referenced articles. In Fig \ref{Others} we plot the above LCs together with their best-fitting model and model residuals. We applied CB-BLS to these systems too but did not detect any significant signal. 

% ========================
\section{Conclusions}
% ========================

We presented a number of improvements to the CB-BLS algorithm that allow CB-BLS to improve it's sensitivity to shallow transits, correctly model distorted close binary stars, use multi-band photometric data, and more. We also presented an analytical approach for the determination of detection limits for CB-BLS. The resultant implementation of CB-BLS was blind-tested using five test light curves derived from CoRoT data --- all five were correctly and strongly identified, and the theoretical prediction for the detection limit was in line with the actual results. We also presented an example of a procedure for the correct detrending of variable stars from wide-field surveys that results in far less systematic input light curves for CB-BLS, and a subsequent first application of CB-BLS to real LCs from both targeted observations and a wide-field survey.

Looking for transiting CB planets requires good understanding of several issues: how to generate good photometry of intrinsically-constant stars (now well understood), how to generate good photometry of intrinsically-variable stars (for e.g., as explained above: iterative detrending of the residuals), high-quality EB modeling, and finally - the actual search for transiting CB planets (e.g., the CB-BLS algorithm). While the above general description will probably remain correct, we are still in the process of learning the details of the process - as exemplified by the still-evolving CB-BLS. Still, we feel that the experience of the single-stars transit surveys allows for an even steeper learning curve (and we remind the reader that single-stars transit surveys had their own learning curve: more than 80\% of the known transiting planets today were not known only three years ago). 

Looking for CB planets around EBs is advantageous since EBs are already well-aligned to our line of sight. Since theoretical models predict that CB planets will be, at least initially, aligned with their host EB -- the chances that CB planets will transit in front of the EB components are relatively high. CB planets in general and transiting CB planets in particular are expected to be particularly beneficial for the general study of extrasolar planets:
\begin{itemize}
\item Binaries are a very significant fraction of the total stellar population. Studying planet formation and evolution without including the formation and evolution of planets in binary star systems is surely incomplete.
\item Frequency of CB planets depends on assumed planetary formation mechanism, and may allow to distinguish between competing theories. The detection limits formalism introduced here will allow us to give meaningful constrains on the frequency of CB planets by correctly accounting for null results.
\item After the initial surprises of close-in giant planets and planets in eccentric orbits, detecting CB planets will again substantially diversify the environments that produce and sustain planets.
\item Understanding very high precision LCs of transiting planets around single stars is limited by the uncertainty of stellar parameters [e.g., Johnson \etal 2008, Gillon \etal 2008, and discussion therein]. EBs allow to us significantly increase stellar parameters accuracy, so resultant fits to planetary structure models will have smaller error bars.
\item Follow up will be interesting for several reasons: I) close-in CB planets can show orbital evolution on short time scales (even 100s of days), II) CB planet systems produce four distinct Rossiter-McLaughlin [e.g., Gaudi \& Winn 2007] effects: 1-2, 2-1, p-1 and p-2 (where 1,2 and p designate the two binary components and the CB planet respectively, and, for e.g., 2-1 stand for the spectroscopic eclipse of component 1 by component 2), III) Higher chances of finding resonant systems and/or chaotic systems.
%\item False positives fraction is also expected to be lower since astrophysical false positives are rare.
%\item EB components may be of very different spectral types ? probe planets in a range of environments with a range of wavelengths.
\end{itemize}

Specifically, using CB-BLS to look for transiting CB planets is especially useful since:
\begin{itemize}
\item CB-BLS allows to find shallow transiting CB planets in the residuals of EBs.
\item CB-BLS can naturally include Roche geometry -- and there are many EBs that require that.
\item CB-BLS can simultaneously fit multi-band data.
\item CB-BLS allows to harness existing datasets to the detection of transiting CB planets with no further inputs. Medium (or maybe even low) resolution spectroscopic input can allow to eliminate the mass ratio as free parameters and to reduce the required CPU time, but this is not mandatory.
\item CB-BLS source code and documentation are planned to be made freely available. The source code is almost entirely in MATLAB and so easily portable.
\end{itemize}

% ========================
\section*{Acknowledgements}
% ========================
We thank Francis O'Donovan and the rest of the TrES team for making the Lyr1 raw data available to us.

AO is supported by the European Helio- and Asteroseismology Network (HELAS), a major international collaboration funded by the European Commission's Sixth Framework Programme.
HD acknowledges support by grant ESP2007-65480-C02-02 of the Spanish Ministerio de Ciencia e Inovaci\'{o}n.

\end{document}